# Investigation of Relevant Distribution System Representation with DG for Voltage Stability Margin Assessment

Alok Kumar Bharati, *Student Member, IEEE*, Venkataramana Ajjarapu, *Fellow, IEEE*

*Abstract*— This paper emphasizes the importance of including the unbalance in the distribution networks for stability studies in power systems. The paper aims to: discuss the various simulation methods for power system analysis; highlight the need for modeling unbalanced distribution system for accurate load margin assessment; demonstrate the influence of net-load unbalance (NLU) on voltage stability margin (VSM). We also share a T&D co-simulation interface with commercial power system solvers. The distribution system is evolving rapidly with high proliferation of distributed energy resources (DERs); these are not guaranteed to proliferate in a balanced manner and uncertainty resulting due to these DERs is well acknowledged. These uncertainties cannot be captured or visualized without representing the distribution system in detail along with the transmission system. We show the impact of proliferation of DERs in various 3-phase proportions on voltage stability margin through T&D co-simulation. We also study the impact of volt/VAR control on voltage stability margin. This analysis is only possible by representing the distribution system in detail through T&D co-simulation. Higher percentage of net-load unbalance (NLU) in distribution system aggravates the stability margin of the distribution system which can further negatively impact the overall stability margin of the system.

## I. Introduction

THE power system is a large complex network of various components that are geographically spread in the form of transmission and distribution sub-systems. The transmission system (T-System) is a large meshed network operating at higher voltages responsible for transmitting bulk electric power from the generating stations to the load centers. The power that is received near the load centers is further distributed to the loads through the power distribution system. The distribution system (D-System) is mostly a radial network operating at lower voltages. The distribution system has been evolving fast with high proliferation of distributed energy resources (DERs) like solar photovoltaic (PV) inverters, smart loads (that can participate in demand response (DR)), electric vehicles (EVs), storage, distributed generation (DG), etc. Traditionally, the power system was designed to handle powerflow in one direction, from the generators to the loads, but, with the inclusion of these DERs in the distribution system there can be two-way powerflow in distribution systems. These changes have occurred mostly over the past few decades and the methods of analyzing power systems has been evolving over these years.

It is important to note that the traditional techniques assume that the distribution system is a passive sub-system of the power system and is generally modeled as a bulk load at the low voltage bus of the transmission network. In many practical system analysis methods, the models used for analysis consider some part of the sub-transmission network but do not consider the distribution system. The bulk power system is mostly balanced in 3-phases, and therefore the power system analysis evolved with per-phase analysis. This enables usage of simpler analytical methods and faster computations. This method of ignoring the distribution system details has been employed for system analysis, operations and controls even at the large control centers for a century and more and the system has been functioning reasonably well as the distribution system was indeed a passive consumer of electric power.

Considering the recent evolution of the power system, the assumptions made in several methods of power system analysis should be re-evaluated. Increasing proliferation of DERs and other smart controllable loads are making the distribution system behavior more uncertain. This uncertainty resulting from the distributed controllable loads and DERs is forcing the inclusion of the distribution system along with the transmission system for more accurate analysis of the power systems. One influence of the high DER penetrations is change in the load compositions observed over the recent years [1]. Phenomena like the duck curve are evidence that the distribution system operating scenarios are changing rapidly.

This paper is submitted for review to IEEE Transactions on Power Systems on 03/12/2019. This work is funded by Power Systems Engineering Research Center (PSERC) project S-70, "Leveraging Conservation Voltage Reduction for Energy Efficiency, Demand Side Control and Voltage Stability Enhancement in Integrated Transmission and Distribution Systems"

Alok Kumar Bharati is a PhD student at the Department of Electrical and Computer Engineering, Iowa State University, Ames, IA 50010 USA (e-mail: alok@ iastate.edu).
Dr. Venkataramana Ajjarapu is a Professor with the Department of Electrical and Computer Engineering, Iowa State University, Ames, IA 50010 USA (e-mail: vajjarap@ iastate.edu).

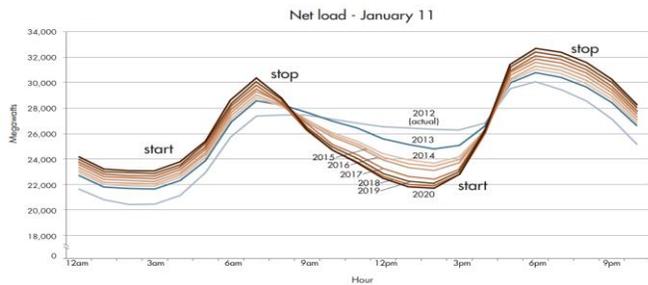

Fig. 1. Extract: California Independent System Operator (CAISO) Duck Curve – 2018[2]

Fig. 1 shows the net load that the ISO must follow for different times of the day and the presence of solar PV in the distribution system is forcing the load to have steeper ramps resulting in the "duck curve". The uncertainty that can arise with natural phenomenon cannot be forecast accurately (as of today) to the extent of incorporating them in the independent system operator (ISO) controls. With these changes, the distribution system is no longer a passive consumer of electric power/ energy, but, in fact, is making the net-load predictions erroneous as described in [3]. These errors are leading to unforeseen operational challenges in the transmission and distribution systems. If these changes in the distribution system are not accounted in the expansion planning studies, it is likely that the system will be over optimistic or pessimistic which are both not good economical and reliable operation of the grid.

The challenge with the detailed modeling of the distribution system for analysis along with the transmission systems is the increase in the simulation time and computational complexity if the entire system is modeled together. The distribution systems should be represented using 3-phase unbalanced representation and if this is done for the entire system, we will lose the benefit of the per-phase analysis for the transmission systems. The simulation time increases exponentially if the entire system is modeled and analyzed using 3-phase unbalanced system in spite of the transmission systems being reasonably balanced. The simulations done, particularly from planning perspectives become very complex and difficult to analyze if the entire system is 3-phase unbalanced. To retain the advantages of the balanced transmission system and simultaneously account for the unbalance in the distribution system, solving them in a decoupled way is ideal. Therefore, T&D co-simulation is an effective method of incorporating the important aspects of transmission and distribution system with sufficient detail.

In the present paper we discuss the various methods of analyzing the power system and understand the impact of the assumptions of each method on the results, especially for the voltage stability studies. The distribution system modeling and the influence of distribution system unbalance will be studied through T&D co-simulation studies. We also show the influence of unbalanced distribution of DG in the three phases of the distribution system. The paper will discuss and demonstrate some concepts using the IEEE 9-bus transmission system, IEEE 4-bus distribution system. At the end, we also show the results and discuss all these aspects using a more detailed model of IEEE 123 bus with all loads modeled as ZIP loads and various distributions of DG in the three phases.

*A. Literature Review*

Several software and analytical tools have been developed to study and understand the power system behavior and perform analysis. It is only over the past couple of decades that the distribution system simulations have gained importance and have been driving lot of recent research using distribution system software like OpenDSS and GridLAB-D [4]. The proliferation of smart and controllable devices is increasing, forcing engineers to study the impact of control of these devices on system behavior. There have been also been analytical methods developed for understanding distribution system stability [5, 6]. All the above-mentioned literature deal with distribution systems in detail but in an isolated way. There has also been some work done to study the impact of distribution system controls on system performance, however they do not focus on the unbalanced nature of the distribution systems [7].

Some of the work assume a balanced distribution system for computational simplicity, while distribution systems are significantly unbalanced in nature [8]. There are some earlier papers that have done some preliminary work on understanding and pointing out the importance and need for transmission and distribution system co-simulation, especially for voltage stability studies [9 -11]. These papers highlight that the two sub-systems (transmission and distribution) can independently drive the system towards voltage instability. The present paper focuses on demonstrating the importance of modeling the distribution system and transmission system in detail. An important part is to acknowledge that the power system community has recognized the need for developing tools to simulate the transmission system and distribution system together [12-15]. However, there is very limited literature available that actually demonstrates the need for T&D co-simulation and the present paper is a contribution in this aspect where we are demonstrating the need for T&D co-simulation and the importance of modeling the transmission and distribution systems together for voltage stability studies.

The North American Electric Reliability Corporation (NERC) is a body regulating the operations and controls in the bulk transmission system and has published multiple reports that states the issues with aggregation of DERs and loads [1]. The contribution of this paper over the state of the art is three fold:

1. Discuss and elaborate the need for modeling of unbalanced distribution system along with the transmission system for voltage stability analysis.
2. Demonstrate the influence of net-load unbalance (NLU) by considering the distributed nature of DERs and loads on the stability margin assessment.

## II. IMPORTANCE OF MODELING THE DISTRIBUTION SYSTEM IN DETAIL TO CAPTURE IMPORTANT PHYSIOGNOMIES

There is a need to understand the importance of considering the distribution system characteristics for overall system studies, e.g., stability limit of distribution system. This section will discuss the importance of modeling the distribution system in detail for complete system studies and is extended to demonstrate the importance specifically for long-term voltage stability margin assessment using P-V curves through T&D co-simulation.

It is important to relate the study to the physical system that operates in reality. In the P-V curve analysis we increase the load on the load bus of the transmission network using a parameter '$\lambda$'. We generally ignore the distribution network that is intermediate to the transmission load bus and the physical locations of the loads that are increased. It is also of interest at this point to understand in detail, which aspects of the system behavior are captured using a method of simulation or analysis and which aspects are totally missed out.

### A. Only Transmission System Simulation (No-D System)

This is the simplest method of power system analysis where the distribution system is modeled as lumped loads. This is mostly followed in academic research and it does not capture any kind of the distribution behavior except for the case where the losses are modeled as a part of the load. This method of modeling the losses is a linear approximation of the distribution feeder loss when used to draw the P-V curve. In reality, the feeder loss is a quadratic function of the load current in that feeder $P_{Loss} \propto I_{Load}^2$.

Apart from the losses, the impact of the voltage drop across the feeder is also not captured in this method of analysis. Capturing the voltage drops along the feeder will impact the study, especially for voltage controls and load control based studies. A fraction of the load is usually voltage dependent load and its value depend on the node voltage and is modeled as ZIP load. ZIP loads are modeled using the following equations:

$$P_{ZIP} = P_0 \left( P_Z \left(\frac{V}{V_0}\right)^2 + P_I \left(\frac{V}{V_0}\right) + P_P \right) \quad (1)$$

$$Q_{ZIP} = Q_0 \left( Q_Z \left(\frac{V}{V_0}\right)^2 + Q_I \left(\frac{V}{V_0}\right) + Q_P \right) \quad (2)$$

Where,
$P_0, Q_0 \rightarrow$ base real and reactive powers of the load
$P_Z, Q_Z \rightarrow$ constant impedance fraction of real & reactive power
$P_I, Q_I \rightarrow$ constant current fractions of real & reactive power
$P_P, Q_P \rightarrow$ constant power fractions of real & reactive power

From Fig. 2, it can be seen that the voltage at the load bus at node 4 is much lower when measured in per unit (pu) than the substation node voltage due to the feeder voltage drop. If the load is modeled as a ZIP load, its value is lower when connected at node 4 than that when connected at node 1.

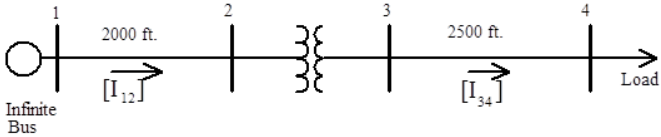

Fig. 2.  IEEE 4-Node Distribution Test Feeder

This approach does not capture the distribution system unbalance which is also an important aspect of the distribution network. Therefore, the next option is to model the distribution network to capture the feeder voltage drop and the losses in a more accurate way. This can be done using an equivalent feeder to represent approximate distribution system.

### B. Transmission System with Equivalent Feeder

The equivalent feeder on the transmission load bus captures the losses in a better way. However, it should be noted that real network has multiple load buses, like in the case shown in Fig. 3, the equivalent distribution feeder (D-Feeder) will not be able to capture the losses accurately as the losses depend on $I_{Load}$ in every feeder segment.

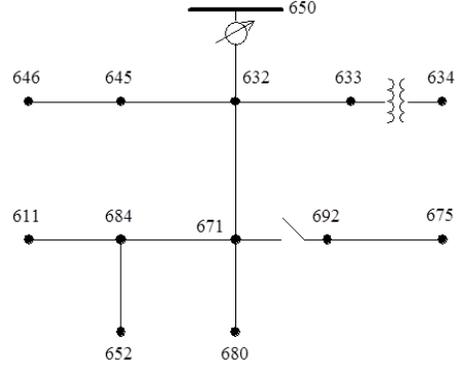

Fig. 3.  IEEE 13-Node Distribution System Test Feeder

Consider there are ZIP loads at all nodes of the network shown in Fig. 3. It can be seen from the figure that there will be a voltage drop from substation or the root node 650 through the multiple feeder segments to the node 652. The load at node 632 will see a voltage that is different than the voltage seen by the load at node 652. This will result in different loads at these nodes which will vary as a function of the node voltage governed by the ZIP load equations (1), (2) mentioned above.

Therefore, due to these different loads at the different operating points, the currents in the feeders are also different resulting in a difference in the total loss. The equivalent feeder cannot capture this effect of varied operating condition and new equivalent feeder parameters need to be computed for every operating point on the P-V curve. However, it is much better than not representing the distribution system at all.

The equivalent feeder parameters are computed based on the net load and net losses in the distribution system and is determined by the following pseudo-algorithm:

1. Determine the substation voltage using transmission system power flow: $V_{sub} \angle \theta_{Sub} = V_{Load} \angle \theta_{Load}$.
2. Perform the distribution system power flow using this substation voltage. We obtain the net power at the substation $P_{Sub}, Q_{Sub}$; $S_{Sub} = \sqrt{P_{Sub}^2 + Q_{Sub}^2}$
3. Identify the net real and reactive power of the load and distribution feeder loss at the substation. Identify the $P_{Load}, Q_{Load}$ and $P_{Loss}, Q_{Loss}$
4. The total net real and reactive power of the distribution system are placed at the transmission load bus. $P_{TransLoad} = P_{Sub}; Q_{TransLoad} = Q_{Sub}$.
5. Calculate the total current flowing into the load modeled (including the losses). $V_L I_L^* = S_{TransLoad} = S_{Sub}$
6. Using the current, calculate the equivalent resistance $R_D$ and equivalent reactance $X_D$ of the equivalent feeder. $I_L^2 (R_D + jX_D) = (P_{Loss} + jQ_{Loss})$

One should note that the $R_D$ and $X_D$ calculated can capture the total distribution system loss, but, it is accurate only for that operating point for which the equivalent feeder parameters are computed. For a different load, different operating point, there will be error in the loss represented by the equivalent feeder and the actual distribution system loss.

The equivalent feeder method obviously does not have any way of representing the unbalance in the distribution system and therefore, the effect of the unbalance in the distribution system is also not captured using the equivalent feeder representation. As mentioned earlier, the equivalent feeder representation can capture the distribution network with good accuracy but, only for an operating point where the equivalent feeder parameters are computed.

### C. The Complete Distribution Network

The distribution system solvers enable us to perform complete 3-phase unbalanced distribution system analysis. These solvers assume the substation or the root node of the distribution system as a swing bus and the voltage is held fixed. This might be a reasonable assumption considering the online load tap changers (OLTCs) at the distribution system, however, it does not take into account the impact of the transmission system voltage. The transmission system is the way by which multiple distribution systems are interconnected and the change in one distribution system will impact the other through the change in the transmission system power flow and the resulting changes in the load bus voltages; especially in voltage stability studies or VSM assessment. Therefore, the only other effective method of capturing both transmission system behavior and distribution system behavior is to co-simulate them.

### D. Transmission and Distribution System Co-Simulation

T&D co-simulation allows for detailed modeling of both, the transmission subsystem and the distribution subsystem. This captures all the details of the distribution system along with the inter-dependent nature of the transmission and the distribution network. However, the tradeoff is that with T&D co-simulation, the computational complexity and computational burden is increased leading to a longer time for simulation but with increased accuracy. For accurately considering both transmission and distribution systems, co-simulation approach is considered in the present research. Distribution system control – grid-edge technologies are fast growing in the distribution networks and are impacting the transmission system operations and analysis. Therefore, modeling the distribution system using the three phase distribution system representation and using the optimized solving methods on the transmission system side can be effectively utilized with T&D co-simulation. However, accurate test systems for T&D co-simulation still need to be established, discussed and debated by the experts in the field who form IEEE working groups.

## III. T&D CO-SIMULATION FRAMEWORK

The need for a transmission and distribution system co-simulation framework using commercial software arises when there is a need to simulate large systems efficiently. T&D co-simulation works on the Master-Slave Splitting (MSS) method of solving coupled subsystems in a de-coupled manner and combine them [12,13]. The MSS is done at the distribution substation node that separates the balanced part of the system and the unbalanced part. The transmission system analysis is done per phase and the distribution system is solved using 3-phase unbalanced system analysis.

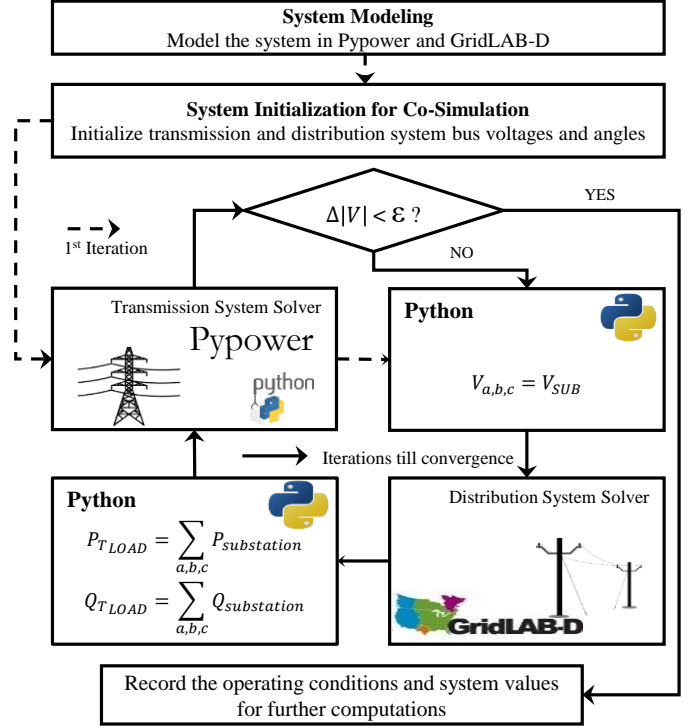

Fig. 4. Illustration of the T&D Co-Simulation is Performed

This method captures the unbalanced nature of the system while preserving the computational ease of per phase analysis. The salient features of the developed co-simulation interface to co-simulate Pypower and GridLAB-D (shown in Fig. 4) are:
- The interface is developed on Python and provides a simple way to operate GridLAB-D which is also open source. [16] provides a base code to write .glm (GridLAB-D model) files using python.
- The code is developed to enable parallel computing for executing multiple T&D co-simulation instances at the required transmission load buses, enabling T&D co-simulation of large systems.
- The interface execution does not need any additional software and has built-in features for plotting and can be extended to generate reports as well.

We have extended the interface to do T&D co-simulation with commercial software on the transmission side and have successfully co-simulated PSSE and GridLAB-D for all steady state analysis.

## IV. SIMULATION CASE STUDIES FOR UNDERSTANDING VSM ASSESSMENT USING T&D CO-SIMULATION

Consider a 2-bus equivalent, but, with an equivalent distribution feeder before the connected load shown in Fig. 5. The 2-bus equivalent shown is modeled with a distribution substation step-down transformer

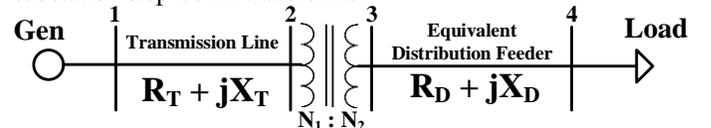

Fig. 5. Extended 2-bus Equivalent with an Equivalent Distribution Feeder

We know for a simple 2-bus equivalent, the voltage stability margin can be determined from the P-V curve derived from the continuation power flow [17]. If the distribution eq. feeder is not represented but the losses are represented as a part of the load, the $R_D$ and $X_D$ values in the extended 2-bus system will be 0 (zero) and this will reduce the effective impedance between the generator and load and the loss is not represented leading to incorrect load margin being estimated.

From the transmission line information, we choose 230 kV line parameters and consider a 320 km line for the transmission line to compute the $R_T$ and $X_T$ in per unit: $R_T = 0.05\ pu;\ X_T = 0.3\ pu$. Let us consider an IEEE 4-bus distribution system parameters to compute the distribution system equivalent feeder parameters. The combined 3-phase load in the IEEE4-bus distribution feeder is 5.4 MW and 2.6153 MW. When a distribution power flow is run for a balanced load on the 4-bus network, the total loss is 0.41938 MW and 0.8614 MVAR. The equivalent feeder parameters calculated assuming the voltage at the load bus is $1\angle 0\ pu$ for the 4-bus distribution network. The equivalent feeder parameters are computed to be $R_D = 0.03\ pu;\ X_D = 0.06\ pu$. The CPF in reference [17] implemented in Matpower is used for drawing the full λ-V curves with and without the distribution feeder (Fig. 6). The $\lambda_{max}$ for the case without eq. feeder is 2.4 whereas it is 1.8 for the case with eq. distribution feeder (D-Feeder).

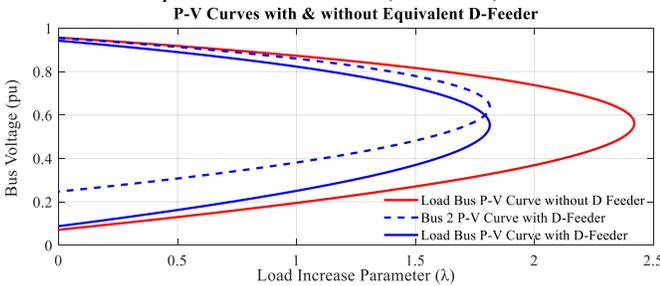

Fig. 6. Voltage Stability Curves for the Extended 2-bus Equivalent System

It is very important to understand the influence of modeling the medium voltage and low voltage networks of the power system while doing power system analysis. Most of the transmission level analysis are done considering the high and the ultra-high voltage networks of the system. With the polymorphism of the distribution networks, it is very important to consider the distribution systems while performing power system analyses, especially analysis related to stability studies. T&D co-simulation is necessary to represent the distribution system accurately and an equivalent distribution feeder cannot capture all the aspects of the distribution system, especially, if it significantly unbalanced. The equivalent feeder parameters are computed with the help of the net load and loss seen at the distribution substation level.

It should be noted that depending on the distribution system and the transmission system modeling the stability margin assessed will be impacted. Accurate estimation of the transmission line parameters and the distribution system equivalents are very important. An important part of the entire power system modeling is the modeling of the sub-transmission network that operate at medium and high voltages. For accurately considering a proposer T&D co-simulation framework, identifying the boundary between balanced and unbalanced part of the system is important.

### A. Importance of Co-Simulating T&D systems for VSM

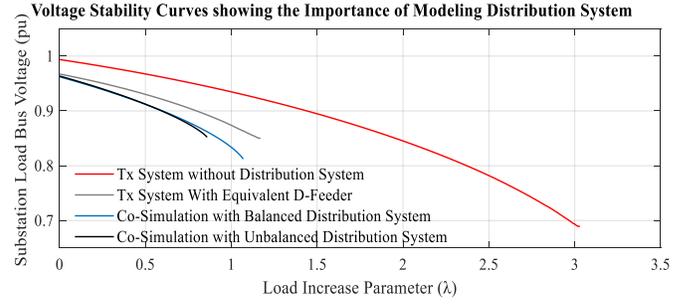

Fig. 7. Voltage Stability Curves with Various Method of Simulations: IEEE 9-Bus Transmission + IEEE 4-Node Distribution System

Fig. 7 shows four curves corresponding to first, second and fourth columns of Table I for IEEE9-bus transmission and IEEE 4-bus distribution system. The distribution feeder is connected at one load bus (125MW) of the transmission system to understand the impact of T&D co-simulation alone.

The red curve corresponds to the transmission system where the losses are modeled as a part of the load i.e., distribution system losses are modeled linearly in the system. If the loss modeling were to be approximately accurate, it results in the grey curve that represents the balanced distribution system almost accurately but, using an equivalent distribution feeder. In case of the equivalent feeder model, the effect of system unbalance is not captured. Therefore, co-simulating transmission and the distribution system together is important to accurately assess the VSM.

The blue and black curves are the voltage stability curves resulting from the transmission and distribution system co-simulation with balanced and unbalanced distribution systems respectively. The equivalent feeder parameters are determined from the net load and losses observed at the substation end. The net 3-phase load and 3-phase losses for the balanced and unbalanced cases are same so the equivalent feeder representing the real and reactive losses are almost the same in both cases. However, the load margin for both these distribution systems are actually different, the unbalanced distribution system reached its stability limit for a smaller increase in load. This cannot be captured using the equivalent feeder model as the equivalent feeder parameters do not depend on the distribution system unbalance. Though the transmission system and the substation are balanced, the individual distribution feeders can be unbalanced, and the next case shows that T&D co-simulation can assess VSM accurately with a balanced substation.

### B. Influence of Unbalanced distribution feeder on VSM through T&D Co-Simulation

It is important to understand that the 3-phase balanced operations that happen in the transmission is due to the aggregated effect of the multiple distribution feeders that are individually unbalanced but when aggregated together are almost balanced out. Much of this balancing is done by design and planning while building the distribution network and part of it is done by active control of balancing the voltages and load either by external devices like voltage regulators, feeder switching, etc. But with the introduction of the recent grid-edge technologies, it has become difficult for the operators to balance the grid and therefore, even if the transmission systems operate at balanced condition, the distribution systems are unbalanced. The distribution system unbalance is usually measured in terms of voltage magnitudes and angles, but, for voltage stability studies, the net load unbalance (NLU) is critical and affects the VSM assessment. The IEEE 4-bus test feeder is modified by adding two more feeders to the substation to make the substation load balanced even if the individual feeders are unbalanced in 3 phases as shown in Fig. 8.

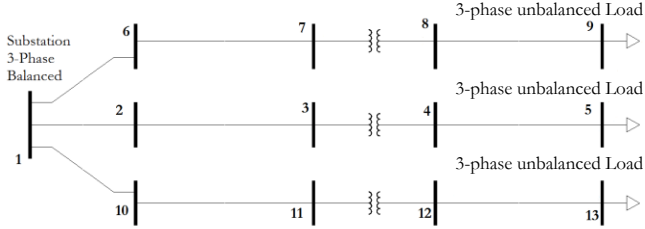

Fig. 8. Modified Distribution System to balance the Substation

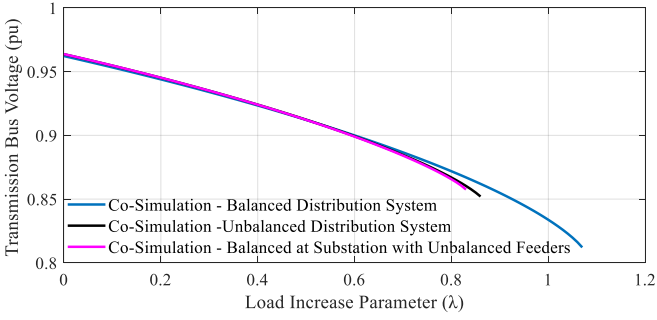

Fig. 9. Voltage Stability Curves for Unbalanced Feeders and Balanced Distribution Systems

Fig. 9 shows the VSM of the modified test system (purple curve) is almost the same as the case with one IEEE 4-bus unbalanced distribution system (black curve). This happens because one of the feeders in the modified system reaches loading limit around the same value of '$\lambda$' as in the case for the IEEE 4- bus case.

### C. Influence of DG Presence on VSM - T&D Co-Simulation

To understand the influence of DG on voltage stability margin of the system, an un-balanced distribution system is considered. Since there is no guarantee that in real world DG penetration is equal on all the three phases, analyzing the influence of DG penetration accurately requires T&D co-simulation. The DG penetration is modeled with various distribution of DG over the three phases to simulate a real-world DG presence. Equivalent (Eq.) feeder method cannot capture the effect of DG being added on some parts of the distribution network as the equivalent feeder parameters are computed based on the 3-phase net load and net losses measured at the substation. But T&D co-simulation can capture the effect of the DG penetration on individual phases as the distribution solver performs a 3-phase unbalanced analysis.

We consider the IEEE 9 bus transmission system and the IEEE 4-bus distribution system for co-simulation and we add DG in different proportions resulting in DG distribution with equal DG proportional to the load on that phase and also in proportions with very low distribution in one phase while the rest is distributed equally in proportion to the load. Fig. 9 shows the PV curves for these cases and also a case with no DG. The DG considered in this case is assumed to be in unity power factor mode (UPF). The load is modeled as ZIP load with profile [ZIP] = [0.4 0.3 0.3]. The distribution system load is unbalanced with the load distribution as: A=28.05 MW; B= 39.6 MW; C=52.25 MW. The total MW of DG added =91.94 MW and distributed in various proportions as a % of load in the three phases (e.g., A=10% means DG is 10% of phase A load).

It can be clearly seen from Fig. 10 that the distribution of DG that results minimum amount of unbalance in the net load has highest load margin, which means the knowledge of the load unbalance and the DG distribution are both important to be reflected in the model of systems that are being studied.

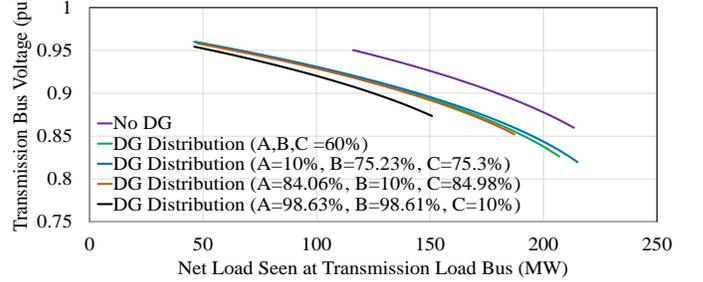

Fig. 10. Voltage Stability Curves for DG Proliferation in various Proportions - IEEE 9-bus Transmission and IEEE 4-bus Distribution Systems

Clearly DG helps increasing the margin but, by how much depends on how the DG is distributed in the system. With the recent amendments to the IEEE 1547 standard, it is important to understand the impact of volt-var control (VVC) on the load margin and how the influence of the unbalance is effected by DG operating in VVC mode.

### D. Impact of Volt-Var Control on VSM through T&D Co-Simulation

Distributed generation, especially the smart inverter based DG are the most prominent among all DG that is proliferating in the distribution system. VVC is essentially controlling the reactive power injection/ absorption based on the node voltage to in-turn control the voltage. We are using a standard VVC with droop control for this analysis. For the same case described in Sub-section C above, we operate the DG in VVC mode and the corresponding results are shown in Table I. We can see from the results that under VVC, the margin is improved compared to the corresponding to unity power factor (UPF) mode. This is because the DG inverters are supplying reactive power and helping the system operate for a larger load transfer before reaching the limit.

TABLE I. IMPACT OF DG OPERATING IN VVC MODE VERSUS UPF MODE ON LOAD MARGIN

| DG Distribution → | %A %B %C | 0, 0, 0 | 60, 60, 60 | 10, 75, 75 | 84, 10, 85 | 99, 98, 10 |
|---|---|---|---|---|---|---|
| → Load Margin/ VSM (MW) | DG in VVC Mode | 97.24 | 171.23 | 185.33 | 145.83 | 108.82 |
| | DG in UPF Mode | 97.24 | 160.79 | 168.27 | 139.33 | 104.74 |

All the results shown so far are for a small distribution system so that it is easier to analyze the results. In the next sub-section,

we simulate the IEEE 123-bus three phase unbalanced distribution system with the IEEE 9-bus transmission system to understand the distribution of DG and the impact of unbalance in the net-load on the voltage stability studies.

*E. Net-Load Unbalance (NLU) and Its Impact on VSM*

We have verified all the above observations using a larger distribution system and have modeled the IEEE 123-bus distribution system along with the IEEE 9–bus transmission system to perform the analysis. The loads are modified to be ZIP loads with ZIP profile [ZIP] = [0.4 0.3 0.3]. The total load on the three phases A, B, C are 45.44 MW, 29.28MW and 36.96 MW respectively. 40%, 60% and 80% DG of the total load is added in various proportions as % of load in the three phases.

Unbalance in distribution system is usually measured in terms of voltage and current magnitudes and angles, however, it is easier to visualize this in the % load unbalance in the three phases, which is explained here. Using the T&D co-simulation, P-V curves are drawn and the load margin is computed. The results are tabulated in Table II. To understand this better, we define the net-load unbalance as percentage by comparing the unbalance with the average net load as below:

$$P_{avg} = \frac{P_A + P_B + P_C}{3} \qquad (3)$$

$$U_i = \frac{P_i - P_{avg}}{P_{avg}} \qquad \forall\ i = A, B, C \qquad (4)$$

$$NLU = \max(abs(U_i)) \times 100\ \% \qquad \forall\ i = A, B, C \qquad (5)$$

Where,
$P_A, P_B, P_C$    The net-loads on phases A, B, C.
$NLU$ %    Percentage of maximum net-load unbalance.

TABLE II. IMPACT OF DG OPERATING IN VVC MODE VERSUS UPF MODE ON LOAD MARGIN

| Total DG Penetration | DG Distribution across 3-Ø (% A,% B,% C) | Net-Load Unbalance (%) | Load Margin/ VSM (MW) | |
|---|---|---|---|---|
| | | | DG in VVC Mode | DG in UPF Mode |
| **40% DG (44.67 MW)** | 50, 10, 50.3 | 17.80 | 273.49 | 243.86 |
| | 40, 40, 40 | 25.53 | 267.69 | 240.61 |
| | 55, 55, 10 | 48.73 | 260.80 | 237.81 |
| | 10, 62, 61 | 88.05 | 241.38 | 231.75 |
| **60% DG (67 MW)** | 60, 60, 60 | 25.53 | 299.17 | 256.44 |
| | 72, 25, 72 | 41.47 | 290.05 | 258.93 |
| | 77, 77, 25 | 85.93 | 284.66 | 254.14 |
| | 25, 85, 85 | 134.85 | 260.10 | 240.98 |
| **80% DG (89.34 MW)** | 90, 50, 90 | 17.80 | 317.48 | 271.28 |
| | 80, 80, 80 | 25.53 | 316.60 | 268.83 |
| | 95, 95, 50 | 147.93 | 296.81 | 269.24 |
| | 52, 100, 100 | 200.00 | 294.25 | 254.48 |

The results for various levels of DG penetration along with the NLU% are shown in the Table II. The results demonstrate all the aspects of the previous sections and sub-sections:

1. The voltage stability margin decreases as the NLU % increases.
2. The NLU% cannot be captured effectively in the equivalent feeder parameters (a more complex equivalent feeder may be necessary to do so).
3. Presence of DG positively impacts the load margin or in other words improves the VSM.
4. DG operating in VVC improves the VSM more compared to DG operating in UPF for any given case.

The balanced or unbalanced distribution of DG is not directly responsible for the impact on VSM, but, the NLU% influences the margin. The DG and flexible load can be controlled to reduce the NLU% and thereby increase the margin.

V. DISCUSSION AND CONCLUSIONS

We have discussed and demonstrated the importance of representing the unbalanced distribution systems in detail using T&D co-simulation methodology. It is understood that with an equivalent distribution feeder, some aspects of the distribution system can be reasonably represented, but, only for balanced systems. The summary of the various methods of representing distribution to capture important physiognomies of distribution systems are described in Table III.

TABLE III. METHODS OF REPRESENTING DISTRIBUTION SYSTEM AND THE TRADE-OFF FOR VSM ASSESSMENT

| Distribution System Physiognomies ↓ | Distribution System Physiognomies Captured ↓ | | | |
|---|---|---|---|---|
| | No D-System | Eq. D-Feeder | Only D System | T&D Co-Simulation |
| **D Losses** | No | Yes (with error) | Yes | Yes |
| **D-Feeder Voltage Drop** | No | Yes (with error) | Yes | Yes |
| **D-Feeder Segment Drop** | No | No | Yes | Yes |
| **Dist. Un-balance** | No | No | Yes | Yes |
| **Impact of T on D** | No | Yes | No | Yes |

It can be seen from Table III that T&D co-simulation can capture the complete power system in the best possible way using the existing solvers. We can still take the advantage of the per phase analysis on the transmission network while capturing the 3-phase unbalance and finer details of the distribution system modeling when we co-simulate transmission and distribution systems.

With the increasing uncertainty of the various DER proliferation, the expected NLU% in the 3-phases of the power network will influence the analysis and simulations. These analyses are the basis of important planning and control decisions and hence should be accurate. These decisions will not only have financial implications but also the control decisions in real time operation may result in operating the grid in less reliable conditions making the grid less secure. The results have not only emphasized the importance of modeling unbalanced distribution networks for voltage stability studies but for understanding the impact of various distribution system controls on the transmission system operations. The smaller test systems used in these studies are developed and accepted by IEEE working groups for transmission and distribution systems and therefore the unbalance in the distribution systems modeled are not arbitrary and should be given due importance. The test systems and the simulations and analyses of the results reflect some important aspects that the power system community need

to consider:
1. T&D Co-simulation is important as it considers the distribution system representation accurately.
2. Unbalance in the distribution systems can influence the analysis results and it needs to be modeled.
3. With the increased DER proliferation in the distribution system, modeling them in full glory is necessary with the detailed distribution system models.
4. Equivalent feeder can only capture minimum representation of the distribution systems but cannot capture the distribution system performances accurately.

*A. Future Work*

There are few points that need to be addressed based on the results and observations from these simulation studies that will impact the T&D co-simulation and also help further to study the importance of modeling the entire system in detail:
1. Accounting for the sub-transmission system and modeling the sub-transmission system accurately.
2. Identifying the boundary bus for T&D co-simulation to separate the system between balanced and unbalanced systems.
3. Development of accurate T&D co-simulation test cases as these will become more relevant very soon and the existing transmission and distribution test systems should be combined for many types of system studies.

The last point is very important in particular because the load levels, the number of parallel feeders, sub-transmission networks and at least the primary distribution feeder modeling will determine the accuracy of the results and analysis. Consider a simple case of IEEE 9 – bus system, the lowest rated load is 90 MW at 345kV. While the highest load on any IEEE distribution feeder is about 5 MW with substation voltage about 12.47kV. We need the intermediate system voltage levels to be accurately represented to analyze the power system with reasonable accuracy. Some larger test-systems have a broader range of voltages covered, but, there is still a gap between the transmission and distribution systems that needs to be filled.

ACKNOWLEDGEMENT

This work is funded from Power System Engineering Research Center (PSERC) project S-70. We also acknowledge the authors of GridLAB-D and Pygrid for providing their open-source software.

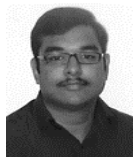
**Alok Kumar Bharati** (S'17) received his Master's degree in Power Electronics and Power Systems from Indian Institute of Technology Hyderabad in 2011. He worked at GE from 2011-17 in R&D of low voltage switchgear. Currently, he is pursuing his Ph.D. in electrical engineering at Iowa State University, Ames, IA, USA.

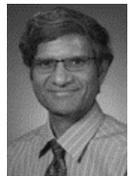
**Venkataramana Ajjarapu** (S'86, M'86, SM'91, F'07) received the Ph.D. degree in electrical engineering from the University of Waterloo, Waterloo, ON, Canada, in 1986.

Currently, he is a Professor in the Department of Electrical and Computer Engineering at Iowa State University, Ames, IA, USA. His present research is in the area of reactive power planning, voltage stability analysis, and nonlinear voltage phenomena.